\documentclass{cernyrep}

\usepackage{hyperref}
\usepackage[varg]{txfonts} 
\usepackage{color}

\newcommand{\erf}{\ensuremath{\,\text{erf}}}   

\newcommand{\arxiv}[1]{\href{http://arXiv.org/abs/#1}{arXiv: #1}}

\title{Statistical methods used in ATLAS for exclusion and discovery}

\author{Diego Casadei
\\
{\small\rm (on behalf of the ATLAS Collaboration)}
}

\institute{Department of Physics, New York University, New York}

\date{\href{http://indico.cern.ch/conferenceDisplay.py?confId=107747}{PHYSTAT
    2011}\\
\today}

\begin{document}

\maketitle

\begin{abstract}
 The statistical methods used by the ATLAS Collaboration for setting
 upper limits or establishing a discovery are reviewed, as they are
 fundamental ingredients in the search for new phenomena.
 The analyses published so far adopted different approaches, choosing
 a frequentist or a Bayesian or a hybrid frequentist-Bayesian method
 to perform a search for new physics and set upper limits.
 In this note, after the introduction of the necessary basic concepts
 of statistical hypothesis testing, a few recommendations are made
 about the preferred approaches to be followed in future analyses.
\end{abstract}

\section{Introduction}

 This note summarizes the statistical methods used so far by the ATLAS
 Collaboration for setting upper limits or establishing a discovery,
 and includes the recommended approaches for future analyses, as
 recently agreed in the context of the ATLAS Statistics Forum.  The
 recommendations aim at achieving a better uniformity across different
 physics analyses and their ultimate goal is to improve the
 sensitivity to new phenomena, while keeping robustness as a
 fundamental request.  The best way to be safe against false
 discoveries is to compare the results obtained using at least two
 different methods,
 at least when one is very near the ``five sigma'' threshold which is
 required in high-energy physics (HEP) to claim a discovery.  One
 recommended method is explained in this paper
 (section~\ref{sec-recomm}).

 We focus here on the searches for some type of ``signal'' in a sample of
 events dominated by other (``background'') physical sources.  The
 events are the output of a particle detector, filtered by
 reconstruction algorithms which compute high-level features like an
 ``electron'' or a ``jet''.  Large use of simulated samples is
 required to tune calibrations, characterize the event reconstruction,
 and compare the outcome of an experiment with the theoretical models.

 A typical simulation consists of few different steps.  First, one
 needs to simulate the result of the primary particle interaction with
 the help of an ``event generator''.  Usually, only one specific
 process of interest is considered (e.g.~Higgs production with a
 specific channel) and saved to disk, allowing the physicist to study
 a well defined type of ``signal''.  Different Monte Carlo (MC)
 productions are then organized to obtain a set of processes which,
 depending on the analysis, can be considered either signal or
 background.  The next step is to simulate the effects of the passage
 of the produced particles (and their decay products) through the
 detector.  This requires the knowledge of the ways energy is
 deposited in each material and defines the ``tracking'' of the
 simulated particles up to the point in which they decay, leave the
 detector or stop.  Finally, the detector response is simulated: for
 each energy deposition into an active material, another MC process
 produces the electronic signal.  The latter is processed in a way
 which closely follows the design of the front-end electronics,
 obtaining the simulated detector output in the same format as the
 data coming from the real detector.

 Statistical uncertainties arise from fluctuations in the energy
 deposition in the active materials and from the electronic noise.
 Systematics due to the limited knowledge of the real detector
 performance and to the details of the offline reconstruction also
 contribute to the final uncertainty and need to be addressed case by
 case.  Finally, theoretical uncertainties in the physical models need
 also to be accounted for.  In general, the differences among the
 event generators cannot be treated as standard deviations, because
 one usually has just two or three available generators.  Hence they
 should not be summed in quadrature but treated separately.

 Section~\ref{sec-notation} summarizes the statistical aspects
 relevant to our problems and defines some notation.  The methods
 applied in past ATLAS publications are reviewed in
 section~\ref{sec-past-methods}, while section~\ref{sec-recomm}
 focuses on the methods which can be used in future analyses.

\section{Notation}\label{sec-notation}

 In HEP we deal with hypothesis testing when making inferences about
 the ``true physical model'': one has to take a decision (e.g.\
 exclusion, discovery) given the experimental data.
 In the classical approach proposed by Fisher, one may decide to
 reject the hypothesis if the \emph{$p$-value}, which is the
 probability of observing a result at least as extreme as the test
 statistic\footnote{A test statistic is a function of the sample which
   is considered as a numerical summary of the data that can be used to
   perform a hypothesis test.} in the assumption that the null
 hypothesis $H_0$ is true, is lower than some threshold.
 In the search for new phenomena, the $p$-value is interpreted as the
 probability to observe at least as many events as the outcome of our
 experiment in the hypothesis of no new physics.
 Alternatively, one may convert the $p$-value into the
 \emph{significance} $Z$, which is the number of Gaussian standard
 deviations which correspond to the same right-tail
 probability\footnote{Here we consider a one-sided test, in which we
   look for an excess over the expected number of events due to the
   background processes.}: $Z = \Phi^{-1}(1-p)$.  The function
 $\Phi^{-1}(x) = \sqrt{2} \erf^{-1}(2x-1)$ is the quantile of the
 normal distribution, expressed in terms of the inverse error
 function.

 A $p$-value threshold of 0.05 corresponds to $Z = 1.64$ and is
 commonly used in HEP for setting upper limits (or one-sided
 confidence limits) with 95\% confidence level.  On the other hand, it
 is customary to require at least a ``five sigma'' level $Z\ge5$
 (i.e.\ $p\le2.87\times10^{-7}$) in order to claim for a discovery of
 a new phenomenon (if $3\le Z \le5$ one usually says only that the
 data suggest the evidence for something new).  It is also common to
 quantify the sensitivity of an experiment by reporting the expected
 significance under the assumption of different hypotheses.

 Another possible approach, suggested by Neyman and Pearson, is to
 compare two alternative hypotheses (if the null hypothesis is the
 main focus of the analysis and no other model is of interest, the
 alternative $H_1$ can be defined as the negation of $H_0$).  In this
 case, two figures of merit are to be taken into account:
\begin{itemize}
 \item the \emph{size}\footnote{$\alpha$ is also known as
         ``significance level'' of the test.  We do not use this
         terminology to avoid confusion with the significance $Z$
         defined above.} $\alpha$ of the test, which is the
       probability of incorrectly rejecting $H_0$ in favour of $H_1$
       when $H_0$ is true.  $\alpha$ is also the false positive (or
       ``type I error'') rate;

 \item the \emph{power} of the test $(1-\beta)$, which is the
       probability of correctly rejecting $H_0$ in favour of $H_1$
       when $H_0$ is false.  $\beta$ is the probability of failing to
       reject a false hypothesis, i.e.\ the false negative (or ``type
       II error'') rate.
\end{itemize}

 In the Bayesian approach, one always compares two (or more) different
 hypotheses.  In order to take the decision among the alternatives,
 one can look at the posterior odds or at the ratio of the marginal
 likelihoods (or ``Bayes' factor'').  The former are always well
 defined and take into account the information accumulated with the
 performed experiment in the light of the existing prior information,
 whereas the latter is often very difficult to compute and may even be
 ill-defined in some problem (for example when comparing two models in
 which one of the priors is improper), although it does not depend on
 the prior knowledge about the hypothesis under consideration.  The
 decision is taken in favour of the hypothesis which maximizes the
 chosen ratio, though the particular value of the latter can can
 suggest a weak, mild or strong preference for that hypothesis.

 In this note, we address two problems, exclusion and discovery, for
 which the notation is different and sometimes misleading, as
 illustrated below.  For this reason, in the rest of the paper we will
 speak about the ``signal plus background'' (sig+bkg or $H_{s+b}$)
 hypothesis and about the ``background only'' (bkg or $H_b$)
 hypothesis, without specifying which is the null hypothesis.

 In the discovery problem, the null hypothesis $H_0$ describes the
 background only, while the alternative $H_1$ describes signal plus
 background.  In the classical approach, one first requires that the
 $p$-value of $H_0$ is found below the given threshold (in HEP one
 requires $p\le2.87\times10^{-7}$).  If this condition is satisfied,
 one looks for an alternative hypothesis which can explain well the
 data\footnote{It might happen that more than a single hypothesis can
   explain the data.  In this case there is no conventionally agreed
   behaviour.  A reasonable approach would be to pick up the one with
   the best agreement with the data, perhaps using a Bayesian
   approach to assess how strong the preference is.}.

 In the exclusion problem, the situation is reversed: $H_0$ describes
 signal plus background while the alternative hypothesis $H_1$
 describes the background only.  In the classical approach, one just
 makes use of the null hypothesis to set the upper limit, though this
 will exclude with probability $\approx\alpha$ parameters values for
 which one has little sensitivity, obtaining ``lucky'' results.
 Historically, this problem has been first addressed in the HEP
 community by the CL$_s$ method \cite{junk99,read02}, whose approach
 is to reject the sig+bkg hypothesis if CLs $= p_{s+b} / (1 - p_b) \le
 \alpha$.  CLs is a ratio of $p$-values which is commonly use in HEP,
 and one can find a probabilistic interpretation if certain asymptotic
 conditions are met \cite{gross2010}.  Another possibility being
 discussed by ATLAS physicists is to construct a Power Constrained
 upper Limit (PCL) by requiring that two conditions hold at the same
 time: (1) the $p$-value is lower than the chosen threshold, and (2)
 the power of the test is larger than a minimum value chosen in
 advance.

\section{Methods used in past ATLAS publications}\label{sec-past-methods}

 So far, different ATLAS analyses used different approaches.
 Converging takes time and is not always possible nor necessarily
 good, the main reason probably being that different uncertainties are
 addressed in different ways.  Whenever possible, the background is
 estimated from data.  Still, one has to extrapolate to the signal
 region and this requires the knowledge of the shape, and hence
 depends on the simulation.  In addition, in many cases signal and
 control regions should be treated at the same time: systematics
 affect both signal and background and often it is impossible to find
 a signal free region.  Finally, in most cases the background is
 composed of several contributions which are independently simulated
 but are not really independent: systematic effects act on all of
 them, making things more and more complicate.

 Accounting simultaneously for systematic effects on different
 components is now possible thanks to HistFactory, a ROOT
 \cite{rootCPC09} tool for a coherent treatment of systematics based
 on RooFit/RooStats \cite{RooStats}, initially developed by K. Cranmer
 and A. Shibata.  First used in the top group \cite{topobs2010},
 HistFactory is now being adopted also by other ATLAS groups.

 Searches for new physics (for example, Higgs searches) often start by
 looking for a ``bump'' in a distribution which is dominated by the
 background.  When the location of the bump is not know, the search is
 typically repeated in different windows, decreasing the sensitivity
 \cite{vitellsLEE,ranucciLEE}.  In the ATLAS dijet resonance search
 \cite{dijet10}, a tool for systematic scans with different methods
 has been applied: BumpHunter, developed G. Choudalakis
 \cite{bumpHunter}.  The program makes a brute force scan for all
 possible bump locations and widths, achieving a very good
 sensitivity, and is appropriate when the bump position and/or width
 are not known. 

 A hybrid Bayesian-frequentist approach has been used by the LEP and
 Tevatron Higgs working groups and is also used in ATLAS Higgs
 searches.  All or some nuisance parameters (modeling systematic
 effects) are treated in the Bayesian way: a prior is defined for each
 parameter which is integrated over.  On the other hand, for the
 parameters of interest the frequentist approach is followed,
 computing $p$-values and constructing confidence intervals.
 HistFactory can be used also with this approach, supporting normal,
 Gamma and log-normal priors for nuisance parameters.

 In the Higgs combination chapter in the ATLAS ``CSC book''
 \cite{atlas-csc-book}, the statistical combination of SM Higgs
 searches in 4 different channels (using MC data) was performed with
 RooFit/RooStats in the frequentist approach: systematics have been
 incorporated by profile likelihood.  Each search was performed with a
 fixed mass and repeated for different values, and the limits have been
 interpolated. Many lessons have been learned and the statistical
 treatment has been refined since then, culminating in the recommended
 frequentist method explained in section~\ref{sec-recomm-freq} below.

\section{Present and future analyses}\label{sec-recomm}

 If possible, one may consider using more than a single approach in
 searches for new phenomena: if they agree, one gains confidence in
 the result; if they disagree, one must understand why (possibly
 finding flaws in the analysis).  This becomes expecially important
 when the obtained sensitivity is close to the minimum limit for
 discovery.  Section~\ref{sec-recomm-freq} below summarizes the
 recently proposed frequentist approach which is being recommended for
 all ATLAS analyses.  A possibility is to test the result of the
 frequentist approach with a Bayesian method.  The current dicussion
 about the Bayesian approach is summarized in
 section~\ref{sec-recomm-bayes}, but at present there is no official
 ATLAS recommendation about it.

\subsection{Recommended frequentist approach}\label{sec-recomm-freq}

 The problem is formulated by stating that the expected number
 of events in bin $i$ is the sum $E(n_i) = \mu s_i + b_i$ of two
 separate contributions, a background expectation of $b_i$ events and
 a signal contribution given by the product of an intensity parameter
 $\mu$ with the expected number of signal events $s_i$.  For
 discovery, we test the background-only hypothesis $\mu=0$.  If there
 is no significant evidence against such hypothesis, we set an upper
 limit on the magnitude of the intensity parameter.

 The Reader will find a full treatment of the recommended method in
 Ref.~\cite{cowan11}.  Very shortly, the profile likelihood is used to
 construct different statistics for testing the alternative bkg and
 sig+bkg hypotheses.  In the asymptotic regime, confidence intervals
 can be found analytically using such statistics, and the resulting
 expressions can be used to define approximate intervals for finite
 samples.  Asymptotically, the maximum likelihood estimate $\hat\mu$
 is Gaussian distributed about the true value with standard deviation
 $\sigma$ which can be found numerically by means of the ``Asimov
 dataset'', defined as the MC sample which, when used to estimate all
 parameters, gives their true values.  In case of exclusion, the
 approximate upper limit (with its uncertainty) is $\hat\mu \pm
 \sigma_{\text{A}} \Phi^{-1}(1-\alpha/2)$.  In case of discovery, in
 which one assumes $\mu=1$, the median significance is
\begin{equation}
  \mathrm{med}[Z_0|1] = \sqrt{2 [ (s+b) \ln(1 + s/b) -s ]}
     \; ,
\end{equation}
 which is the recommended formula for a counting experiment by the
 ATLAS Statistics Forum when estimating the sensitivity for discovery
 \cite{cowan11}.

\subsection{Current discussions about the Bayesian approach}\label{sec-recomm-bayes}

 In the Bayesian approach, the full solution to an inference problem
 about the ``true physical model'', which is responsible for the
 outcome of an experiment, is provided by the posterior probability
 distribution of the parameter of interest.  Typically, there are
 several nuisance parameters which model systematic effects or
 uninteresting degrees of freedom.  In order to obtain the marginal
 posterior probability distribution as a function only of the
 parameter of interest, one has to integrate over all nuisance
 parameters.  This marginalization procedure contrasts with the
 frequentist approach based on the profile likelihood, in which the
 nuisance parameters are fixed at their ``best'' values.

 Prior probabilities need to be specified for all parameters and
 should model our knowledge about the effects which they refer to.
 Quite often, one does not want to encode a precise model into the
 prior or does not assume any relevant prior information.  In this
 case, uniform densities are commonly preferred for computational
 reasons, but they are often misinterpreted as ``non-informative''
 priors, which is not the case.  For example, a uniform density is no
 more flat, when considered as a function of the logarithm of the
 given parameter.  When attempting to make an ``objective'' inference,
 least-informative priors should be used instead.  They can be
 defined, as in the case of the reference priors, as as the ones which
 maximize the amount of missing information \cite{berger2009}.
 Reference priors are invariant under reparametrization, are known
 (and often identical to Jeffreys' priors) for most common
 one-dimensional problems in HEP, and can also be used to test the
 dependence of the result from the choice of the prior
 \cite{bernardo2009,bernardo2011,demortier2011}.

 When dealing with discovery or exclusion in the Bayesian approach,
 one has to make a choice between two alternative hypotheses:
 background only ($H_b$) and signal plus background ($H_{s+b}$).
 Comparing the posterior probabilities is the best way to account for
 the whole amount of information provided by the experiment in the
 light of the previous knowledge.  Although values of $O(1000)$ for
 the posterior odds are interpreted as a strong preference, no
 widespread agreement exists in the HEP community about a minimum
 threshold for claiming a discovery.\footnote{A possible approach
   could be to simulate many pseudo-experiments, compute the $p$-value
   and follow the ``five sigma'' rule mentioned above.}
 In order to check the impact of the assumptions made before
 performing the experiment on the final decision, it is also useful to
 compare the posterior odds against the prior odds (defined as the
 ratio of prior probabilities for $H_b$ and $H_{s+b}$, whenever this
 is well defined).

\section{Summary}\label{sec-summary}

 This note summarizes the statistical approaches used in the past
 ATLAS analyses and the current ongoing efforts to provide uniformity
 of statistical treatment across all analyses.  Guidelines for
 estimating the sensitivity with a frequentist method based on profile
 likelihood ratio have been recently formalized \cite{cowan11}.  In
 this approach, which is recommended for all ATLAS analyses, all
 nuisance parameters are fixed at their best values and a single MC
 sample (the Asimov dataset) can be used to find the numerical values
 of the interesting statistics.

 The Bayesian approach can also be considered in the analysis,
 although no official ATLAS recommendation has been made yet about the
 best method.  In general, the prior densities should be chosen in the
 way which best models our prior knowledge of the model.  Whenever one
 wants to minimize the impact of the choice of the prior on the
 result, one should be aware that flat priors are to be considered
 informative.  On the other hand, least-informative priors can be
 defined for all common HEP problems and have very appealing
 properties.  In the Bayesian approach, the treatment of systematics
 is different from the recommended frequentist method, because the
 whole range of each nuisance parameter is considered in the
 marginalization.  Hence, the comparison between the two approaches
 may be helpful, expecially near the sensitivity threshold for
 discovery.


\end{document}